# Interference Modeling in Cognitive Radio Networks: A Survey

Mohsen Riahi Manesh and Naima Kaabouch
[1]Department of Electrical Engineering, University of North Dakota, Grand Forks, ND 58202, USA
[+]mohsen.riahimanesh@und.edu

**ABSTRACT**

One of the fundamental elements impacting the performance of a wireless system is *interference*, which has been a long-term issue in wireless networks. In the case of cognitive radio (CR) networks, the problem of interference is tremendously crucial. In other words, CR keeps the important promise of not producing any harmful interference to the primary user (PU) system. Thus, it is essential to investigate the impact of interference caused to the PUs so that its detrimental effect on the performance of the PU system performance is reduced. Study of cognitive interference generally includes developing a model to statistically demonstrate the power of cognitive interference at the PUs, which then can be utilized to examine different performance measures. Having inspected the different models for channel interference present in the literature, it can be obviously seen that interference models have been gradually evolved in terms of complication and sophistication. Although numerous papers can be found in the literature that have investigated different models for interference, to the authors' best knowledge, very few publications are available that provide a review of all models and their comparisons. This paper is a collection of state-of-the-art in interference modeling which overviews and compares different models in the literature to provide the valuable insights for researchers when modeling the interference in a specific scenario.

## 1      INTRODUCTION

One of the most widely used technologies that have dramatically impacted the living styles of the people is wireless communications. The transmission of a signal wirelessly across the Atlantic Ocean by Marconi in 1901 verified the realism of wireless communication and opened a new age of technology. Since the last two decades, wireless communications have been involved in every part of human lives, ranging from highly available Wi-Fi hotspots to hardly realized underwater acoustic communication to deep space communications, from distributed ad hoc-based systems to centralized infrastructure-oriented systems, and from greatly commercialized TV broadcasting to confidentially military radio systems. Nevertheless, demands for wireless applications are still growing. Traditionally, governing organizations such as Federal Communication Commission (FCC) statically allocate spectrum channels to entities. As a result, the problem of scarcity of the spectrum is arising as higher number of devices go wireless. In the meantime, investigations show that a huge fraction of the radio spectrum is underutilized in time, frequency, and space [1], [2]. For instance, it is shown in [3] and [4] that about 85% of some spectrum bands in the range of less than 3 GHz are temporally wasted. The disparity between reasonable spectrum scarcity and underutilization of the spectrum has motivated the use of so-called dynamic spectrum access (DSA) method, which suggests the spectrum to be dynamically shared and reused [5], [6]. The proposed technology to realize and enable the DSA is cognitive radio. A CR is able to detect and scan its adjacent environment and dynamically and adaptively adjust its functional parameters to coexist with the licensed (primary) systems in a non-interfering way [1]. CR is intended to be an intelligent solution to prominently enhance the spectrum usage by opportunistically exploiting the underutilized portion of the frequency bands possessed by licensed systems [7], [8].

One of the fundamental factors affecting the performance of wireless systems is *interference*, which has been a long-term issue in wireless networks. It is not an overstatement at all to point out that wireless communication is just fighting impairment and interference of wireless channels. In the case of CR networks, the problem of interference is tremendously crucial. In other words, CR keeps the important promise of not producing any harmful interference to the primary system. Figure 1 provides an idea of how detrimental the effect of interference on CR networks is. As it can be seen from the figure, bit error probability (BEP) tends to dramatically decrease when there is no interference. However, with the increase of the interference power, signal to interference ratio (SIR) decreases which leads to a substantial increase in BEP. Thus, it is essential to realize and investigate the interference caused to the PUs so that its detrimental effect on the performance of the primary system is reduced.

Study of cognitive interference generally includes developing a model to statistically demonstrate the power of cognitive interference at the PUs, which then can be utilized to examine different performance measures. Having inspected the different models for channel interference present in the literature, it can be obviously seen that interference models have gradually evolved in terms of complication and sophistication. Beginning from the naivest models such as the *co-located* network proposed in [9], other models based on constant interference and communication ranges, as well as those models trying to model "capture" (such as, the capture threshold model employed in the network simulator *ns2* [10]) have been suggested and employed. Gupta and Kumar [11] proposed the *physical* or *SINR model* as a more accurate tool to model the interference contrary to the capture threshold model that imprecisely compares the desired signal strength with merely one interfering signal each time. In this manner, if the signal-to-interference-plus-noise-ratio (SINR) observed by the receiver surpasses a SINR-threshold for the whole period of the transmission, a wireless transmission of signal is said to succeed. The SINR takes the cumulative interference power into consideration, and is a more regular metric for determining interference, than the ratio of the desired signal strength to separate interfering signal strength as in the capture threshold model.

In this paper, we give an overview of different interference models and research issues in modeling the interference in cognitive radio. Provided the difficulty of the topic and the variety of existing models, our work is by no means exhaustive. We believe that this paper will provide a preview of the current researches in modeling the interference in the cognitive radio networks. In the rest of the paper, the terms aggregate, additive and accumulative interference are interchangeable.

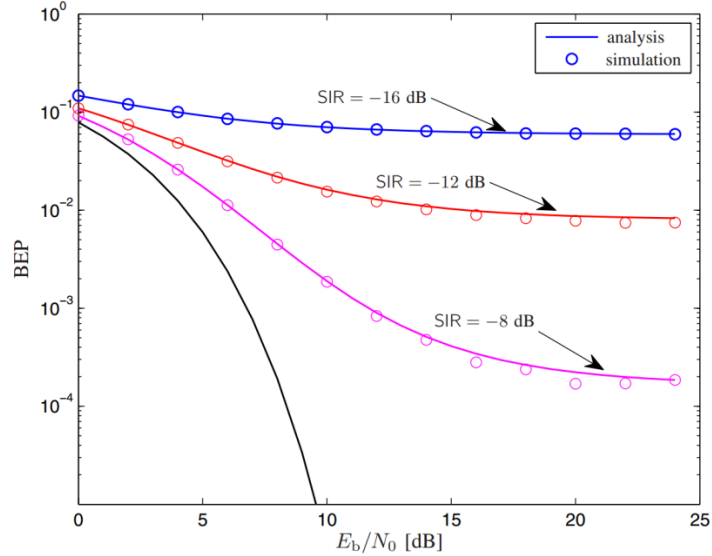

Fig 1. Bit error probability (BEP) of a BPSK system in the presence of cognitive network interference [12]

## 2  MODELS FOR WIRELESS CHANNEL INTERFERENCE

As mentioned in the introduction, different models with different complexity have been proposed and employed for wireless interference in the literature. However, there is, in fact, no agreement on the terminology for wireless interference models in the literature. Therefore, to be accurate in our discussion, we specifically present the wireless channel interference models that we desire to examine. As shown in Figure 2, interference modeling can be classified in two categories: spatial modeling and SINR modeling. The first category includes two types: Protocol Model and Interference Range Model. The second category is also divided in two types: Aggregate Interference Model and Capture Threshold Model.

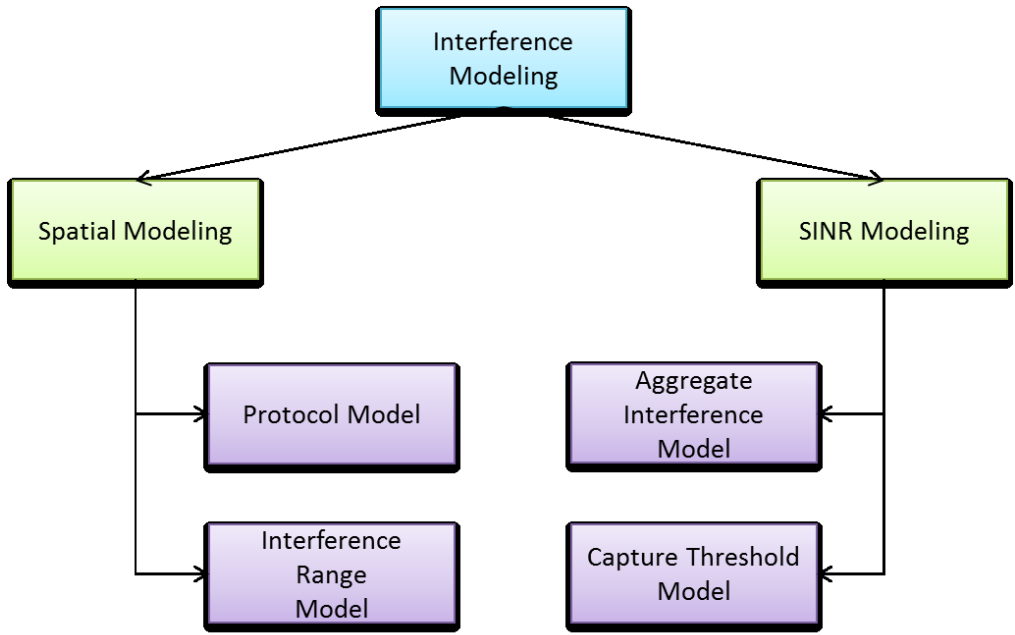

Fig. 2. Classification of different interference models

## 2.1 SPATIAL-BASED MODELING

For spatial-based interference modelling, as indicated by the term, interference changes in accordance with the geographical situation and the separation of the transmitters and the receivers in the network. For example, in [13] and [14], the portion of accessible white holes for CR networks was studied. In [15], the CR receivers' interference zone and CR transmitters' communication zone were examined for the case where a CR network collocates with a cellular network. The coexistence of CR network with wireless microphones functioning in TV bands was investigated in [16], in which the lack of robust communication region because of the presence of CR devices was analyzed. In particular, spatial based interference model is divided into two categories of protocol model and interference range model that are discussed in the following.

### 2.1.1 PROTOCOL MODEL

The so-called protocol model [11], also known as unified disk graph model, has been broadly employed by scientists and investigators in wireless networking society as a means to ease the mathematical description of physical layer. Based on the protocol model, a signal transmission succeeds when a receiver is placed in the transmission region of its wanted transmitter and is located out of the interference region of other unwanted transmitters. According to the this model, the influence of interference from a transmitter is binary and is merely specified by whether or not a receiver goes inside the interference region of this transmitter. In other words, as shown in Figure 3, if a node is inside the interference region of an unwanted transmitter, it is assumed to get the interference and; therefore, it cannot properly receive from its preferred transmitter; else, the interference is considered to be trivial. Hence, the protocol model has been usually employed in evolving protocols and methods in wireless networks because of its simplicity, and can be straightforwardly applied to investigate large-scale wireless systems [17-24].

The argument about the protocol model is that a binary judgment of existence or non-existence of the interference does not precisely provide physical layer features. For example, the protocol model considers that a node will not receive appropriately from its desired transmitter once the node is located within the interference region of an unwanted transmitter. However, this is exaggeratedly pessimistic, as some successful transmissions might still happen even with the presence interference based on capacity formula. On the other side, protocol model indicates that no interference will take place if a node is outside of the interference region of every undesired transmitter. Nevertheless, this idea is rather too optimistic, as slight interference from various sources can accumulate which might not be deniable in calculating the capacity of the channel. As a result, there have been some thoughtful uncertainties in the research society on the accuracy and perfection of protocol interference model employed in wireless networks.

Based on the protocol model [11], a successful data transmission on link $S$ takes place, if for each link $P \in U \setminus \{S\}$ the following inequality satisfies.

$$D(P_T, S_R) \geq (1 + \rho)D(P_T, P_R) \text{ and } D(S_T, S_R) \leq R_c \qquad (1)$$

where $\rho$ is the spatial protection margin which is a positive parameter greater than 0 and $R_c$ stands for communication range. $P_T$ and $S_T$ denote the transmitters and $P_R$ and $S_R$ represents the receiver of the links *P* and *S*, respectively. *U* denotes the set of simultaneously active links. *D(X ,Y)* represents the distance between nodes *X* and *Y*.

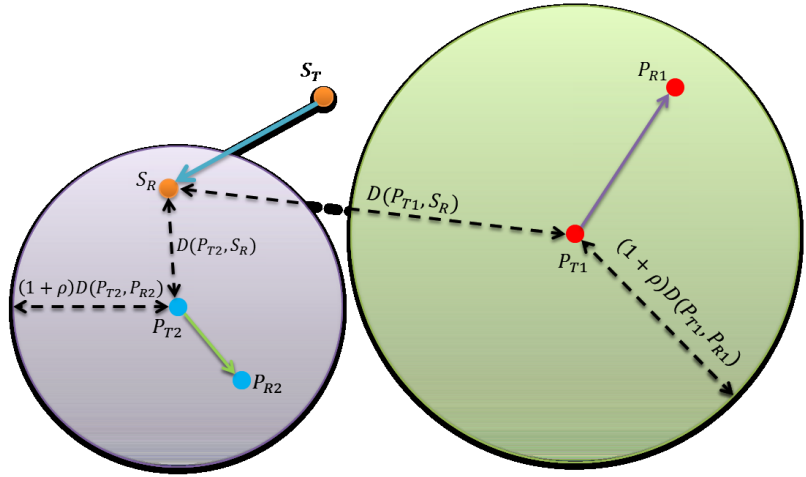

Fig. 3. Based on the protocol model $P_{T2}$ interferes with $S_R$. However, since $S_R$ is out of interference range of $P_{T1}$, no interference happens

### 2.1.2 INTERFERENCE RANGE MODEL

The interference range model is a simplified version of protocol model and takes into account constant ranges for communication and interference [17, 25]. In other words, under the *interference range* model, any node inside a *specific* geographical distance from a receiver is expected to interfere while according to the *protocol model* [11], there is a distance-based relationship between the desired sender-receiver pair and any possible interferer. Specifically, based on the interference range model, a successful packet reception on link *S* happens, if for each link $P \in U \setminus \{S\}$

$$D(S_T, P_R) \geq R_I \text{ and } D(S_T, S_R) \leq R_c \qquad (2)$$

in which $R_I$ stands for interference range, and $R_c$ stands for communication range.

### 2.2 SINR-BASED MODELING

The basis of the SINR model, also known as the physical model, is transceiver design of communication systems that considers interference as noise. Regarding the SINR model, a successful transmission happens provided that SINR at the desired receiver surpasses a threshold to facilitate the transmitted signal to be decoded with a satisfactory bit error probability. Moreover, capacity calculation is based on SINR (via Shannon's formula), which takes into account interference due to concurrent transmissions by other nodes. Since practical coding methods exist in this model to reach its solution in real systems, such interference model is known as a reference model in wireless communications. Nevertheless, the problem of the physical model is that it is difficult to gain a solution due to its computational complexity, mostly when it involves cross-layer optimization in a multi-hop network environment. The reason is that based on the transmission powers, SINR is a non-convex function. Consequently, it is difficult to develop a solution to cross-layer optimization by means of the physical model, and it could cause computational complexity for large-scale networks. As a result, most of the existing methods for cross-layer optimization using the physical layer model stick to a basic layer-by-layer (or "layer-decoupled") approach and therefore come up with sub-optimal solutions [26-28] or as an alternative, revolve around preparing asymptotic lower and upper bounds [11], [29], [30].

### 2.2.1 AGGREGATE INTERFERENCE MODEL

In order to support the notion of opportunistic spectrum sharing, the CRs are required to verify that they do not cause too much interference to the licensed users. It will be very difficult to persuade primary users to coexist with CRs if there is no guarantee about the interference. However, in the context of wireless communications, it is really challenging to give promises to the primary users on the level of interference. This point has been mentioned in [31], [32], where large margins were presented in the sensing threshold requirements to be responsible for undeterminable fading and shadowing. Even after considering these margins, too much interference may still happen by a single CR which meets its personal sensing restrictions when that CR concurrently transmits with another CR meeting its sensing limitations. This is considered as aggregate interference. Even though a private district of radius $L$, as shown in Figure 4, is specified where no CR is permitted to operate, the CRs' additive interference is destructive to the general performance of the system.

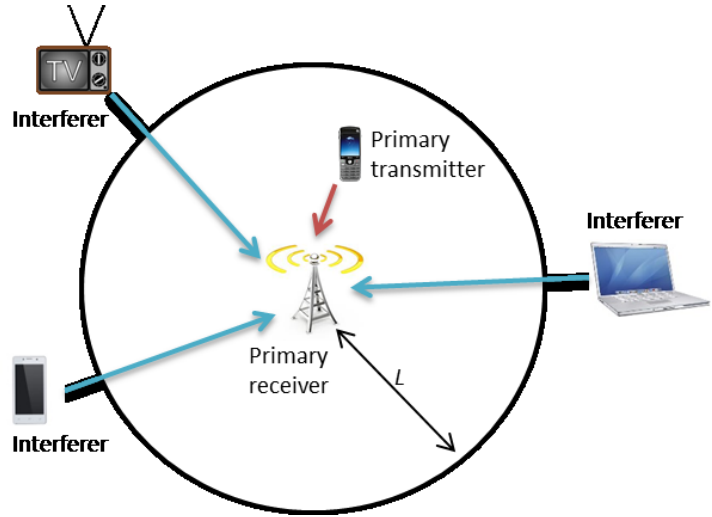

Fig. 4. Aggregate interference

Indicating the transmit power utilized by the transmitter of link $S$ as $Y_S$, the SINR observed by the receiver of link $P$ is provided by:

$$SINR_{P_R} = \frac{G_{P_T \to P_R} Y_P}{(\sum_{S \in U} G_{S_T \to P_R} Y_S) + Y_N} \quad (3)$$

where $G_{X \to Y}$ denotes the channel gain from the point $X$ to the point $Y$, and $Y_N$ denotes the thermal noise power in the frequency band of interest. The addition in the denominator is happened over all links $S \in U \setminus \{P\}$ where $U$ represents the set of all simultaneously active transmissions. It is said a packet reception is successful (or interference is tolerable) if all over the period of the transmission:

$$SINR_{P_R} \geq th_P \tag{4}$$

where $SINR_{P_R}$ is calculated by (3), and $th_P$ is an SINR threshold equivalent to an acceptable bit error probability (BEP), which depends on the modulation and coding methods employed by link *P* [33]. Therefore, based on the additive interference model, if at any time throughout the transmission, equation (4) does not satisfy, there will be a packet loss.

When modeling the SUs' aggregate interference on PU, it is vital to take into account the subsequent factors and features of CRN [34]:

- The distribution style of nodes in the location of the interest such as Poisson point process (PPP) [35-37], Binomial point process (BPP) [38], and uniform distribution [39], [40];
- The geographic dispersion shape of SU node such as exclusion regions [41], [42], circular regions [41-43], and non-circular regions [36], [44];
- The propagation features of the wireless channel, such as path loss, shadowing, and multipath fading [45-47].
- The spectrum scanning methods/PU detection techniques such as energy detection [45], [48], matched filter detection [49], Euclidian distance [50], and sending beacon signal [35], [51], [45];
- The cooperation among SUs;
- The transmission parameters of nodes, such as transmission power, frequency band and modulation.

The probability density function (PDF) of the aggregate interference and the outage probability of a primary receiver due to the interference are two commonly used statistics for the aggregate interference modelling [52], [53].

The PPP has been broadly employed in analysis of interference because of its appropriateness to describe the distribution of nodes in wireless communications [54]. The dispersion shape of nodes geographically is also essential for examination of the interference. Some works [41], [42], [55] assume exclusion area around PU in which SUs are not permitted to enter in the region, though it is normally complicated to assume such an area for passive PU receivers. Some patterns have limited or boundless regions of circular or non-circular shape. Spectrum sensing and PU detection is the significant element of cognitive radio, which should be taken into consideration in modelling the interference. The outage probability or PDF of interference employed in the modeling procedure expressively differs for different spectrum sensing techniques. Analysis of interference for wireless communications turns out to be more complicated due to the complexity of the propagation environment. The probability model of the interference has been assessed in the literature for only path loss [56], path loss and shadowing [57], path loss plus shadowing and Rayleigh fading [51], [54], [58] or Rician Fading [51], [45] and Nakagami-*m* fading [41], [45], [59], [60]. Log-normal shadowing incorporated in co-channel interference was also considered in [61].

Figure 5 shows the PDFs of the aggregate interference power with Poisson parameter $\lambda = 1$ and different values of the exclusive region radius *L*. When *L* is sufficiently large, it is found that the interference at a primary receiver can be approximated by a confined Gaussian-like distribution, a much more desirable distribution for secondary spectrum access systems compared with a heavy-tailed a-stable distribution. Figure 5 provides an insight that the disruptive aggregate interference is mainly caused by a small number of dominant interferers nearby the victim receiver. Once these dominant interferers are eliminated using an exclusive region, the aggregate interference power can be reduced significantly. It should be noted that, given certain constraints on the aggregate interference power, eliminating a few dominant secondary interferers within the exclusive region can allow more secondary users outside the exclusive region to transmit. In other words, the overall secondary traffic may increase if no secondary transmission is allowed within the exclusive region.

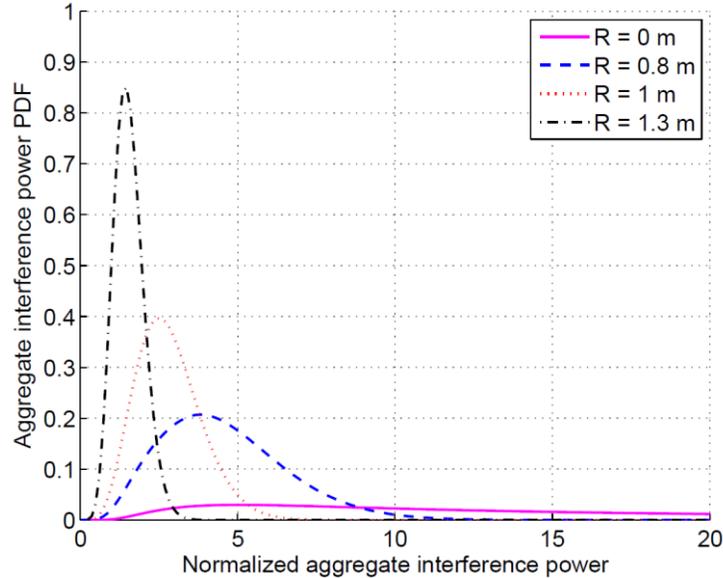

Fig. 5. PDFs of the aggregate interference power with Poisson parameter λ=1 and different values of the exclusive region radius $L$ [62]

### 2.2.2 CAPTURE THRESHOLD MODEL

This model was simulated using NS-2 network simulator [10] and can be considered as a simplified version of aggregate interference model. It does not take into account the interference power of all active users accumulatively, but individually. In other words, the interference is accounted for only one interferer at a time. Specifically, there is no interference on a link $P$ from a link $S$ provided that,

$$G_{P_T \rightarrow P_R} \Upsilon_P \geq \tau_P \quad (5)$$

and

$$\frac{G_{P_T \rightarrow P_R} \Upsilon_P}{G_{S_T \rightarrow P_R} \Upsilon_S} \geq \theta_P \; for \; all \; S \in U \quad (6)$$

where $\tau_P$ and $\theta_P$ denotes the receive threshold and capture threshold, respectively. As it can be seen, the interference is accounted for only one interferer at a time. The desired received signal power is compared with the receive threshold. If it is smaller than the receive threshold the received signal is discarded. However, if it is higher than the receive threshold, the ratio of the received signal power to each individual interfering signal power (SINR) is computed and compared with the capture threshold. If even one interfering link can be found that (6) does not satisfy, the desired received packet is discarded and interference is assumed to have happened.

## 3    COMPARISON OF INTERFERENCE MODELS

As shown in Figure 6 [63], when the transmitter and the receiver are close enough, the minimum range of interference anticipated by the capture threshold model and the one under the additive (aggregate) interference model are fairly similar. Yet, as the transmitter-receiver distance becomes larger, it is no longer precise to estimate the minimum interference range as a factor of the transmitter-receiver separation. Especially, the minimum range of interference can be subjectively high for very "weak" links (*i.e.*, when the signal can be hardly decoded from the transmitter). The interference range model is obviously simple using a fixed value of interference range for every transmitter-receiver separation.

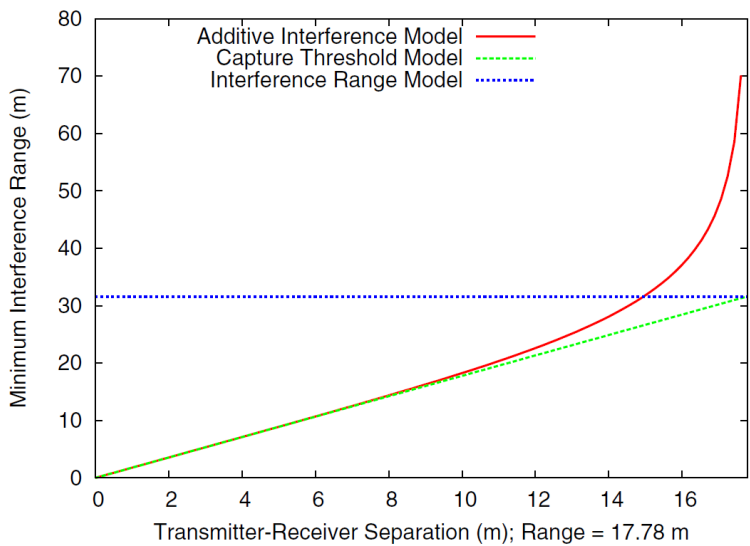

Fig. 6. Minimum interference range predicted by different interference models [63]

The way that different models use to conduct the interference power is the most important reason for this difference in the field of minimum interference range. In the additive interference model, the major quantity of interest is the SINR, through which the preferred signal strength can be compared to the addition of the noise and additive interference powers. In the capture threshold model, the strength of the desired signal is independently compared to the receive threshold $\tau_P$ and to the individual interference terms. This results in a minimum range of interference that is only dependent on the transmitter-receiver distance for the capture threshold model, but on the other hand, as the sum of the noise power and all the interference terms are used in the additive interference model, the minimum interference range increases sharply close to the communication range.

The type of conflict relationships is another significant difference between the interference models presented formerly. In other words, there is a *pair-wise* or *binary* relationship between the wireless links in the capture threshold, the protocol and the interference range models; that is, any two links either interfere together, or they can be working at the same time, irrespective of the other continuing signal transmissions. Conversely, although each link in a set of links may not interfere with a certain link in the additive interference model, interference could still happen. Accordingly, it is not correct to refer to the conflict relationships between just a link *l* and a link *m* in isolation. Rather, as presented in [12], it is likely to compute the conflict sets of link *l* which are all probable sets of links and which could cause a transmission on link *l* to fail by being active at the same time. The reason is that the only model (among those we studied) that behaves interference additively is the aggregate interference model. Significant ways can be drawn from the differences in the conflict relationships between the wireless links, to predict capacity, performance and so on. However, the main problem with this model is that no particular feature is provided to develop effective decentralized algorithms. Table 1 summarizes the pros and cons of the two different interference models of SINR based and spatial based modeling.

Table 1. Comparison between SINR- and spatial-based interference models

|  | PROS | CONS |
|---|---|---|
| **SINR Model** <br> **(Aggregate interference)** | • Can handle features of wireless radio propagation such as Rayleigh fading and shadowing effects. <br> • Graded interference <br> • Accurate | • Does not provide any specific feature to develop efficient decentralized algorithms <br> • High computationally complex |
| **Spatial Model** <br> **(Protocol model)** | • Easy to use <br> • Low computationally complex | • Pair-wise (Binary interference) <br> • Inaccurate |

## 4 CONCLUSION

One of the fundamental elements affecting the performance of a wireless system is interference, which has been a long-term issue in wireless networks. In the situation of CR networks, the problem of interference is tremendously crucial. Several types of multi-hop wireless communication systems, such as cognitive networks, do not provide similar amount of centralized control. However, they further involve in a distributed resource assignment. In such systems, interference is not closely manageable and is exposed to significant uncertainty which makes it difficult to model. In addition, another challenge is to model the interference in CR network with a low complexity while considering the parameters such as the distribution of the nodes over the area, PU detection or the spectrum sensing method, the cooperation between nodes, the channel and transmission properties and etc. Moreover, the impact of a hidden primary receiver and/or hidden SU transmitters on the perceived interference must also be considered. In this paper, different interference models were investigated. It was shown that the interference range model and protocol model, which depend on the transmitter-receiver separation, are the simplest ones yet most inaccurate. Capture threshold and aggregate model are more accurate depending on SINR of the incoming signals. However, since aggregate model takes all incoming signals into account, it is a more complex technique. The most important difference between spatial- and SINR-based models is in the manner they deal with interference. Pair-wise assumption of spatial-based models puts them at the end of the selection list. Our discussions will hopefully motivate future works to develop more realistic interference models.